\documentclass[useAMS,usenatbib,aas_macros,fleqn]{mn2e}

\usepackage{epsfig}
\usepackage{amssymb}
\usepackage{natbib}
\usepackage{aas_macros}
\usepackage{times}
\usepackage{color}
 \usepackage[normalem]{ulem}

\usepackage{amsmath}
\usepackage[normalem]{ulem}

\DeclareGraphicsExtensions{.eps}

\newcommand{\figu}{Figure~}
\newcommand{\figus}{Figures~}
\newcommand{\eq}{Equation~}

\newcommand{\sect}{Section~}

\newcommand{\ergse}{{\rm erg\, s^{-1}}}

\newcommand{\sis}{$\sigma$}
\newcommand{\sise}{\sigma}

\newcommand{\mbh}{$M_{\rm bh}$}
\newcommand{\mbhe}{M_{\rm bh}}

\newcommand{\mstar}{$M_{\rm star}$}
\newcommand{\mstare}{M_{\rm star}}

\newcommand{\msune}{M_{\odot}}

\begin{document}

\def\sarc{$^{\prime\prime}\!\!.$}
\def\arcsec{$^{\prime\prime}$}
\def\arcmin{$^{\prime}$}
\def\degr{$^{\circ}$}
\def\seco{$^{\rm s}\!\!.$}
\def\ls{\lower 2pt \hbox{$\;\scriptscriptstyle \buildrel<\over\sim\;$}}
\def\gs{\lower 2pt \hbox{$\;\scriptscriptstyle \buildrel>\over\sim\;$}}

\title[SMBHs: selection bias and models]{Selection bias in dynamically-measured supermassive black hole samples: Scaling relations and correlations between residuals in semi-analytic galaxy formation models}

\author[E. Barausse et al.]
{Enrico Barausse$^{1}$\thanks{E-mail:$\;$barausse@iap.fr},
Francesco Shankar$^{2}$\thanks{E-mail:$\;$F.Shankar@soton.ac.uk},
Mariangela Bernardi$^{3}$, Yohan Dubois$^{1}$ \newauthor
and Ravi K. Sheth$^{3}$
\\
$1$ Institut d'Astrophysique de Paris, UMR 7095, CNRS \& UPMC, Sorbonne Universit\'{e}s, 98bis Bd Arago, F-75014 Paris, France\\
$2$ Department of Physics and Astronomy, University of Southampton, Highfield, SO17 1BJ, UK\\
$3$ Department of Physics and Astronomy, University of Pennsylvania, 209 South 33rd St, Philadelphia, PA 19104\\
}
\date{}
\pagerange{\pageref{firstpage}--
\pageref{lastpage}} \pubyear{2017}
\maketitle
\label{firstpage}

\begin{abstract}
Recent work has confirmed that the scaling relations between the
masses of supermassive black holes and host-galaxy properties such as
stellar masses and
velocity dispersions may be biased high. Much of this may be caused by
the requirement that the black-hole sphere of influence must be
resolved for the black-hole mass to be reliably estimated. We revisit
this issue with a comprehensive galaxy evolution semi-analytic model.
Once tuned to reproduce the (mean) correlation of black-hole mass with
velocity dispersion, the model cannot account for the correlation with
stellar mass.  This is independent of the model's parameters,
thus suggesting an internal inconsistency in the data.  The predicted
distributions, especially at the low-mass end, are also much broader
than observed.  However, if selection effects are included, the
model's predictions tend to align with the observations.
We also demonstrate that the correlations between the \emph{residuals}
of the scaling relations are more effective than the relations
themselves at constraining AGN feedback models. In fact, we find that
our model, while in apparent broad agreement with the scaling
relations when accounting for selection biases, yields very weak
correlations between their residuals at fixed stellar mass, in stark
contrast with observations. This problem persists when changing the
AGN feedback strength, and is also present in the hydrodynamic
cosmological simulation Horizon-AGN, which includes state-of-the-art
treatments of AGN feedback. This suggests that current AGN feedback
models are too weak or simply not capturing the effect of the black
hole on the stellar velocity dispersion.
\end{abstract}

\begin{keywords}
(galaxies:) quasars: supermassive black holes -- galaxies: fundamental parameters -- galaxies: nuclei -- galaxies: structure -- black hole physics
\end{keywords}

\section{Introduction}
\label{sec|intro}

Supermassive black holes, with masses $\mbhe\sim 10^6 $--$10^9 \msune$, have been identified at the centres of all local galaxies observed with high enough sensitivity \citep[see, e.g.,][for reviews]{FerrareseFord,ShankarReview,KormendyHo,GrahamReview15,2013ApJ...764..184M}.  A surprising finding that has puzzled astrophysicists for the last forty years or so is that the masses of these black holes appear to be tightly linked to the global properties of their hosts, such as stellar mass or velocity dispersion, defined on scales up to a thousand times the sphere of influence of the central black hole.  The origin of these correlations is still hotly debated, though there is general agreement that understanding this origin will shed light on the more general and still unsolved problem of the formation of galaxies.

Supermassive black holes are thought to have formed in a highly star-forming, gas-rich phase at early cosmological epochs. Central ``seed'' black holes are thought to gradually grow via mainly gas accretion, eventually becoming massive enough to shine as quasars or Seyfert galaxies and trigger powerful winds and/or jets capable of removing gas and quenching star formation in the host galaxy. This feedback from active black holes has become a key ingredient in almost all galaxy evolution models \citep[e.g.,][]{Granato04,Ciras05,Vittorini05,Croton06,Hopkins06,Lapi06,Shankar06,Monaco07,Guo11,Barausse12, dubois12a,dubois16,FF15,Bower16}. At later times, both the host galaxy and its black hole may further increase their mass (and size) via mergers with other galaxies/black holes, which could contribute up to $\sim 80\%$ of their final mass \citep[e.g.,][]{DeLucia07,Malbon07,oser10,Shankar10,oser12,Gonzalez11,Shankar13,dubois13,dubois16,Rod16,welker17}. Additional mechanisms, besides mergers, can also contribute to the growth of the stellar bulge and feeding of the central black hole, most notably disc instabilities \citep[e.g.,][]{Bower06,Bournaud11a,DiMatteo12,Barausse12,dubois12b}.

Recently, \citet[][see also \citealt{Bernardi07}, \citealt{gultekin}, \citealt{morabito} and \citealt{Remco15}]{Shankar16BH} showed that the local sample of galaxies with dynamical mass measurements of supermassive black holes is biased. Local galaxies hosting supermassive black holes, irrespective of their exact morphological type or the aperture within which the velocity dispersion is measured, typically present velocity dispersions that are substantially larger than those of a very large and unbiased sample from the Sloan Digital Sky Survey (SDSS) with similar stellar masses. One of the main reasons for this bias can be traced back to the observationally imposed requirement that the black-hole gravitational sphere of influence must be resolved for the black-hole mass to be reliably estimated. Via dedicated Monte Carlo simulations and accurate analysis of the residuals around the observed black-hole scaling relations, \citet{Shankar16BH} found the velocity dispersion to be a more fundamental quantity than stellar mass or effective radius. Indeed, the observed black-hole scaling
relation involving the stellar mass was found to be much more biased than the one involving velocity dispersion (up to an order of magnitude in normalisation), and its apparent tightness could be entirely ascribed to a selection effect.

\citet{Shankar16BH} also suggested that a selection bias more prominent in stellar mass than in velocity dispersion may explain several discrepancies often reported in the literature, \textit{i.e.} the fact that the observed relation between black-hole and stellar mass predicts a local black-hole mass density two to three times higher than inferred from the relation between black-hole mass and velocity dispersion \citep[e.g.,][]{Graham07BHMF,Tundo07,SWM}. \citet{shankar_new} further extended the comparison between the set of local galaxies with dynamically measured black-hole masses and SDSS galaxies. They found evidence that even the correlation between black-hole mass and S\'{e}rsic index, recently claimed to be even tighter than the one with velocity dispersion \citep{Savo16n}, is severely biased, with the correlation to velocity dispersion remaining more fundamental. The bias in the local scaling relations could also have profound implications for the background of gravitational waves expected from binary supermassive black holes, which could be a factor of a few lower than what current pulsar timing arrays can effectively detect \citep[][]{Sesana16}.

The aim of the present work is to revisit the local scaling relations between the masses of supermassive black holes and host-galaxy properties, namely velocity dispersion and stellar mass, in the context of a comprehensive semi-analytic model of galaxy formation and evolution  \citep[][]{Barausse12}. This model also evolves supermassive black holes self-consistently from high-redshift ``seeds'', and accounts for black-hole mergers and for the feedback from active galactic nuclei (AGNs). After briefly reviewing the model in \sect\ref{sec|model}, we discuss (Sections \ref{sec:normalisation} and \ref{sec:dispersion}) the slope, normalisation and scatter of the black-hole scaling relations with and without the aforementioned selection effect on the resolvability of the black-hole sphere of influence.
In Section \ref{sec:residuals} we study the correlations between the residuals from fitted scaling relations, and show that they are useful for constraining theoretical models such as ours as well as
the hydrodynamic cosmological simulation Horizon-AGN \citep{dubois,dubois16,2016MNRAS.460.2979V}. Finally, we will discuss our results in \sect\ref{sec|discu} and summarise our conclusions in \sect\ref{sec|conclu}.

\begin{figure*}
    \center{\includegraphics[width=14truecm]{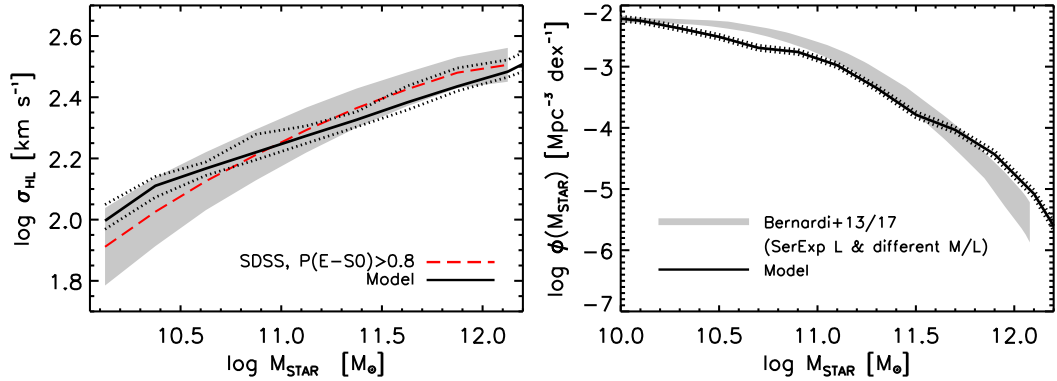}
    \caption{Left: Velocity dispersion as a function of total stellar mass for SDSS galaxies with $P(E+S0)>0.8$ (long-dashed red line), with its 1$\sigma$ dispersion (grey), compared with the prediction of the light-seed model for bulge-dominated/elliptical galaxies, i.e. galaxies with $B/T>0.7$ (solid line, with dotted lines marking the 70\% confidence region). The model's predictions for heavy seeds are very similar. Right:  Stellar mass function of SDSS galaxies based on S\'{e}rsic plus Exponential fits to the observed surface brightness (shaded area, accounting for the uncertainties due to the stellar population modelling, fitting, and assumptions about dust in the galaxies, c.f. \citealt{Bernardi13,Bernardi17}). Solid and dotted lines show the prediction of the model with light seeds and its (Poissonian) $1\sigma$ uncertainties. The predictions for heavy seeds are very similar.
    \label{fig|ScalingsMstar}}}
\end{figure*}
\section{Model}
\label{sec|model}

The full description of the semi-analytic model adopted as a reference in this work can be found in \citet{Barausse12}, with later updates
of the prescriptions for the black-hole spin and nuclear star cluster evolution described respectively in \citet{Sesana14} and \citet{Antonini15,Antonini15b}. Here, we briefly summarise the key points about the growth of the central supermassive black holes, which are the main focus of this paper.

The model is built on top of dark-matter merger trees generated via extended Press-Schechter algorithms \citep[e.g.,][]{Press74,Parki08} tuned to reproduce the results of N-body simulations~\citep{Parki08}. Galaxies form in each halo via the interplay and balance of gas cooling, star formation and (supernova) feedback. Dark matter haloes are also initially seeded with either \emph{light} black holes of $M_{\rm seed}\sim200\, \msune$ (to be interpreted \textit{e.g.} as the remnants of PopIII stars), or with \emph{heavy} black holes of mass $M_{\rm seed}\sim 10^5\, \msune$, which may arise for instance from protogalactic disc instabilities. The seeding of haloes is assumed to happen at early epochs $z>15$, with halo occupation fractions depending on the specific seeding model \citep[see][for details]{Barausse12,Klein16}.

In our model, seed black holes initially grow via (mainly) gas accretion from a gas reservoir, which
is in turn assumed to form at a rate proportional to the bulge star formation rate \citep[e.g.,][]{Granato04,Lapi06}. As a result, the feeding of this reservoir and the ensuing black-hole accretion events typically happen after star formation bursts triggered by major galactic mergers and disc instabilities.
In both their radiatively efficient (``quasar'') and inefficient (``radio'') accretion modes, the black holes also exert a feedback on the host galaxies,
thus reducing their (hot and cold) gas content and quenching star formation. As discussed by a number of groups \citep[][]{Granato04,Ciras05}, AGN feedback prescriptions such as these tend to induce a correlation between black hole mass and velocity dispersion of the bulge component. Also accounted by the model is the black-hole growth via black-hole mergers, following the coalescence of the host galaxies. This mechanism becomes particularly important for high black-hole masses at recent epochs.

\begin{figure*}
    \center{\includegraphics[width=14truecm]{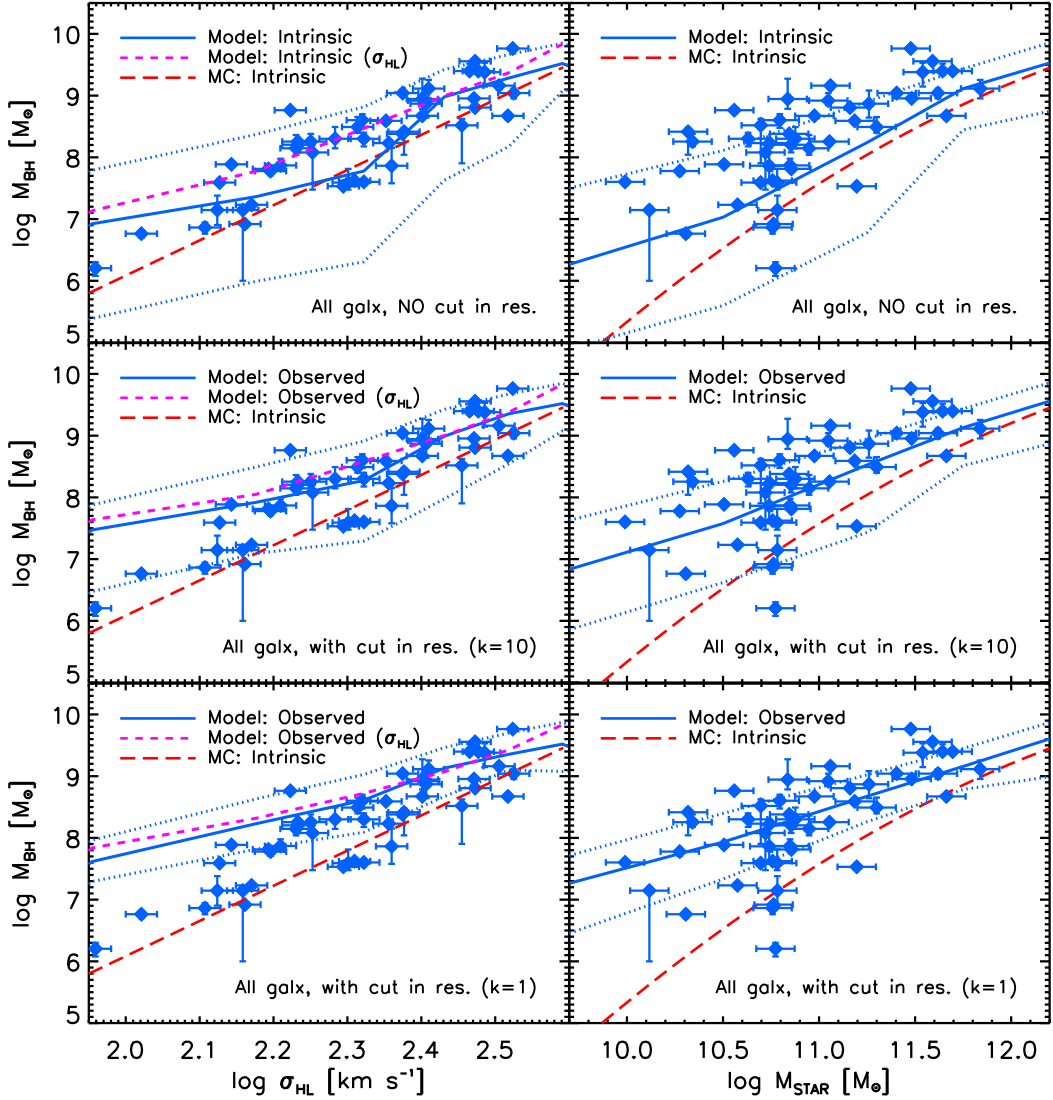}
    \caption{Black-hole mass as a function of velocity dispersion (left panels) and total stellar mass (right panels) as predicted by our model with light seeds (the results for heavy seeds are qualitatively similar). Results are shown for the full outputs of the model (labelled as ``Intrinsic'', top panels), and for the subsample of black holes with a gravitational sphere of influence above 0.1'' (labelled as ``Observed'', middle and bottom panels; see text for details). The solid and dotted blue lines in all panels represent the medians and 70\% confidence region of the distributions. The dashed magenta lines in the left panels are the median model predictions when assigning velocity dispersions to galaxies from the observed SDSS $\sigma_{\rm HL}-\mstare$ relation (long-dashed red line in the left panel of \figu\ref{fig|ScalingsMstar}). Long-dashed red lines are the intrinsic scaling relations as inferred from Monte Carlo simulations by \citet{Shankar16BH}. Blue diamonds are data collected and updated by \citet{Savo15} on local galaxies with dynamical measurements of supermassive black holes. Note that observational biases can increase the normalisation and reduce the scatter in the intrinsic scaling relations.
    \label{fig|ScalingsLight}}}
\end{figure*}

The model is calibrated against a set of observables, such as the local  stellar and black-hole mass functions, the local gas fraction, the star-formation history, the AGN luminosity function,
the local morphological fractions, and the correlations between black holes and galaxies and between black holes and nuclear star clusters~\citep[c.f. ][]{Barausse12,Sesana14,Antonini15,Antonini15b}. In more detail, as we will show in the following (c.f. Fig.~\ref{fig|ScalingsLight}), the
model's default calibration attempts to match the observed \mbh-\sis\ relation without accounting for any observational bias (on morphological type  or on the resolvability of the black-hole influence sphere).

\section{Results}
\label{sec|results}

We will now compare the predictions of our model with observations, focusing on the normalisation of the scaling relations and on the role played by selection biases; the dispersion around the scaling relations; and the correlations between the residuals of the data from the scaling relations.

\subsection{The normalisation of the scaling relations and the role of selection bias}
\label{sec:normalisation}
The left panel of \figu\ref{fig|ScalingsMstar} shows the relation between velocity dispersion and stellar mass for
early-type galaxies in the SDSS.  Here `early-type' means that the probability of being elliptical or lenticular, $p$(E+S0), according to the automatic morphological classification of \citet{Huertas11}, exceeds 0.8.  We restrict to this specific SDSS subsample as velocity dispersions in late-type galaxies are not spatially resolved, though the correlation does not depend on the exact cut in $p$(E+S0).  For consistency with the data of \citet{Savo15} to which we will compare, we follow \citet{Shankar16BH} and correct the velocity dispersions $\sigma_{\rm HL}$, as in \citet{Cappellari06}, to a common aperture of 0.595 kpc \citep[\textit{i.e.} the one adopted by the Hyperleda data base,][]{Paturel03}. Henceforth, unless stated otherwise, we will always define velocity dispersions \sis\ at the aperture of Hyperleda.  Stellar masses \mstar\ are from \citet{Bernardi13}.  They are the product of luminosity $L$ and mass-to-light ratio $\mstare/L$; the $L$ values are from \citet{2015MNRAS.446.3943M}, based on S\'{e}rsic+Exponential fits to the light profiles.

The black solid line marks the median velocity dispersion-stellar mass relation as predicted by the model (for bulge-dominated/elliptical galaxies only) and black dotted lines show the 15th and 85th percentiles of the predicted distribution (at fixed stellar mass). Central velocity dispersions in the model are computed as $\sigma=A \sqrt{G M_{\rm b}/{r_{\rm b}}} [1+(V_{\rm b}/\sigma)^2]$,
where $M_{\rm b}$ is the bulge dynamical mass, $r_{\rm b}$ is the scale radius of the Hernquist profile  \citep[which the model adopts to describe the bulge, see][]{Barausse12}, $A\approx 0.4$ accounts for the anisotropy of the distribution function of the bulge stellar population \citep[c.f.][\figu2, lower panel]{Baes02},
and the ratio $V_{\rm b}/\sigma$ accounts for the contribution of the bulge rotation and is modeled based on observations \citep[c.f.][for details]{Sesana14}.
As can be seen, the predicted correlation is similar to the observed one, although slightly flatter.

\begin{figure*}
    \center{\includegraphics[width=14truecm]{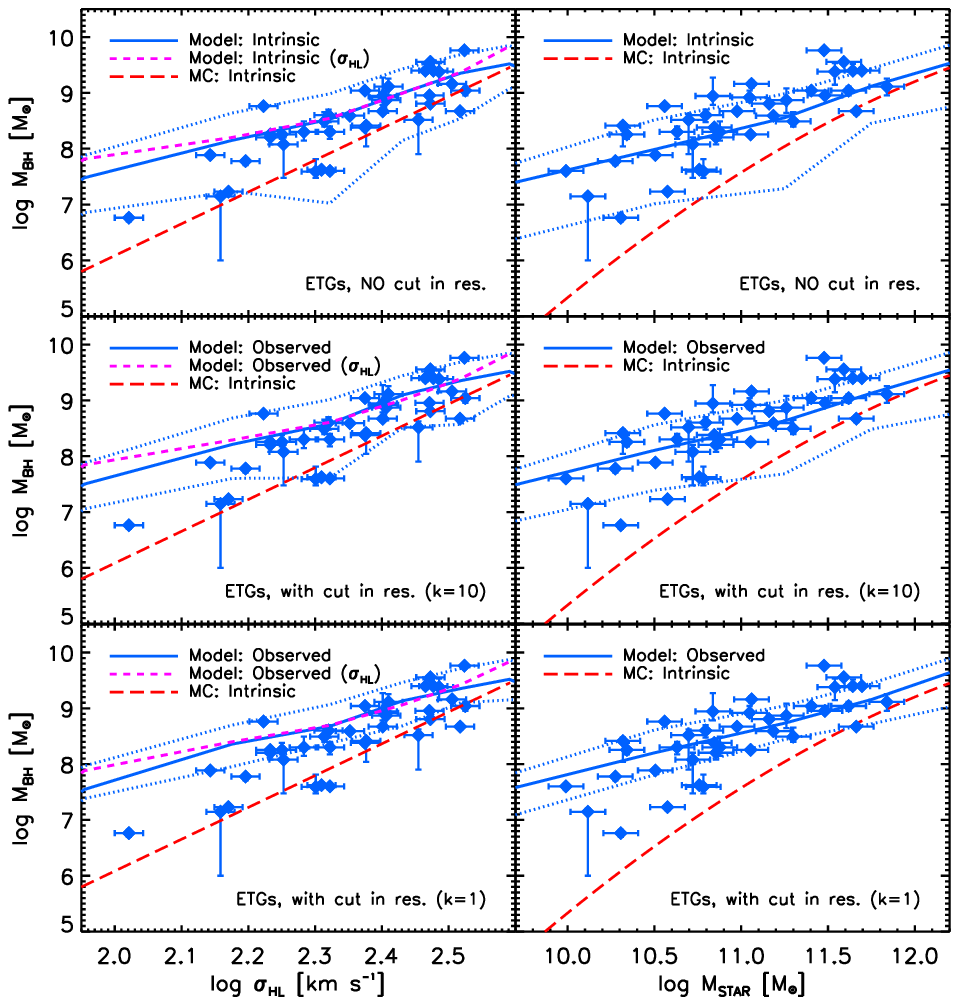}
    \caption{Same as \figu\ref{fig|ScalingsLight}, but for galaxies with a bulge-to-total ratio $B/T>0.7$ (and for light seeds; the results are similar for heavy seeds). The results are broadly similar to \figu\ref{fig|ScalingsLight}, though the dispersion of the model's intrinsic predictions is substantially lower than for the full galaxy sample. Note that the observational dataset is also restricted to early-type galaxies only, to match the sample of simulated galaxies.
    \label{fig|ScalingsETGs}}}
\end{figure*}

For completeness, the right panel of \figu\ref{fig|ScalingsMstar} compares the stellar mass function predicted by the model with the observed one \citep{Bernardi13,Bernardi17}.
While the model lies slightly above the data at the highest masses, it lies below over the range $2\times 10^{10} \lesssim \mstare/\msune \lesssim 10^{11}$. This is not a major issue in the present context, since the model is consistent with the empirical galaxy scaling relations, and most notably with the \sis-\mstar\ relation shown in the left panel.

Having checked that our model reproduces the dynamical scaling relations of early-type galaxies, we now study the scaling relations with the central supermassive black hole. \figu\ref{fig|ScalingsLight} compares the model (with no morphological selection) with the whole sample (i.e., spirals as well as ellipticals and lenticulars) of \citet[][blue diamonds]{Savo15}.  The left and right panels show the scaling of black hole mass with velocity dispersion and total stellar mass, respectively. In each panel, the solid and dotted lines show the median and the region containing 70\% of the model objects in a given bin. The top panels show the full black-hole sample, while the middle and bottom panels only show the subset for which the sphere of influence exceeds the typical (HST) resolution limit,
\begin{equation}
 r_{\rm infl}\equiv k\frac{G\mbhe}{\sigma^2}\,, \qquad \frac{r_{\rm infl}}{d_{\rm Ang}}>0.1''\,
 \label{eq|rinfl}
\end{equation}
($d_{\rm Ang}$ being the angular-diameter distance).
We use the parameter $k$ to take into account different galaxy mass profiles:  $k\sim 4$ for the Hernquist profiles assumed by the model \citep[see][for details]{Barausse12}, but $k\sim 10$ (or even larger) is possible if a core is present.
On the other hand, strong lensing and accurate dynamical modelling have shown that the mass profiles of intermediate-mass, early-type galaxies are consistent with nearly isothermal profiles down to (at least) tenths of the effective radius \citep[e.g.,][and references therein]{Cappellari15}.  These have $k\sim 1$. To bracket these uncertainties, we show results for both $k=10$ and 1.

To match the data as closely as possible, we draw the angular diameter distances $d_{\rm Ang}$ from an empirical probability distribution function,
which we construct from the distances in the \citet{Savo15} sample, and which peaks around $\sim 15$--20 Mpc.  Using a distribution that is uniform in comoving volume, as done in \citet{Shankar16BH}, yields comparable results.

The top panels of \figu\ref{fig|ScalingsLight} suggest that the model predicts intrinsic scaling relations that lie slightly below the data, especially for \mbh-\mstar\ (top right), but which are broadly consistent with the intrinsic relations suggested by \citet[][long dashed red lines]{Shankar16BH}.  The match to the \mbh-\sis\ relation improves (slightly) if we assign velocity dispersions by using the SDSS $\sigma$-$\mstare$ relation, e.g. via the analytic fits provided by
\citet[][dashed magenta line, left panels]{Sesana16}.\footnote{Note that it makes sense to assign \sis\
from the model-predicted \mstar\ via the
observed $\sigma$-$\mstare$ relation, rather
than \mstar\ from the model-predicted \sis, because masses
are more ``primitive'' quantities than velocity dispersions for
a semi-analytic galaxy formation model such as ours.}
  At the same time, the model substantially underpredicts the \mbh-\mstar\ relation by up to an order of magnitude (top right panel), suggesting that there is some internal inconsistency in the data. In other words, a model like ours, tuned to match the local velocity dispersion-stellar mass and black-hole mass-velocity dispersion relations, tends to severely underpredict the \mbh-\mstar\ relation.  This is in line with the results of \citet[][]{Shankar16BH}. Calibrating the model to match the observed \mbh-\mstar\ relation would instead overestimate the observed \mbh-\sis\ relation.  Such an overestimate has indeed been seen in two (very different) cosmological hydrodynamic simulations \citep{Sija15,2016MNRAS.460.2979V}.\footnote{\citet{2016MNRAS.460.2979V} mention resolution, which is at best 1 kpc in their simulation, as one reason why their results do not match the \mbh-\sis\ relation. However, their predictions for \mbh\ are larger than the observations even at large \sis\ (c.f. their \figu7), while they are in good agreement with the \mbh-\mstar\ relation.}

  When the selection effect on the sphere of influence of the black hole is  applied to the model (middle and bottom panels of \figu\ref{fig|ScalingsLight}), the median normalisations of the predicted scaling relations increase (especially for the \mbh-\mstar\ relation), because a substantial fraction of low-mass black holes are excluded.  This is because, for a given angular aperture, \eq\ref{eq|rinfl} preferentially removes objects with the smallest gravitational spheres of influence; these tend to be the lowest-mass black holes.
   Therefore, this effect tends to select the ``upper end'' (in black-hole mass) of the intrinsic distributions shown in the top panels of \figu\ref{fig|ScalingsLight}. This also induces an overall flattening of the scaling relations, which is again more obvious in the \mbh-\mstar\ plane: selection hardly matters for the most massive galaxies, but it causes a factor $\lesssim 10$ increase in the median observed \mbh\ at lower masses. Selection-biased models are flatter in the \mbh-\sis\ plane as well.\footnote{As a result of this flatter slope, the model's prediction (after applying the selection bias)
   lies above the (few) data with $\sigma\sim 10^2$ km/s in the sample of \citet{Savo15}. Note however that other samples, such as that of
\citet{2011Natur.469..374K}, contain black holes with masses up to $\sim 10^8 M_\odot$ at $\sigma\sim 10^2$ km/s, which is in better agreement with our model.}

We conclude that, to agree with resolution-biased observed scaling relations, models must predict intrinsic scaling relations that are significantly \emph{steeper} than the observed relations.
We have tried to obtain steeper intrinsic scaling relations in our model by changing the AGN feedback, but we have found this to be insufficient to achieve better agreement with the data at low masses and velocity dispersions. In fact, the results and conclusions of this paper are robust to changes in the AGN feedback strength as well as to changes in the black-hole accretion prescriptions.

\figu\ref{fig|ScalingsLight} also clearly shows that in addition to changing the normalisation and slope, selection dramatically decreases the dispersion around the median relations, especially at lower masses.
Note that these ``corrections'' do not depend on the exact choice of $k$, since they are present for both $k=10$ and $k=1$.

Similar comments apply to \figu\ref{fig|ScalingsETGs}, which compares model-predicted galaxies with bulge-to-total ratios of $B/T>0.7$ with the E+S0 galaxies of \citet{Savo15}, in the same format as \figu\ref{fig|ScalingsLight}.  The intrinsic distributions of the model-predicted galaxies (top panels) are narrower than in the full samples (c.f. top panels of \figu\ref{fig|ScalingsLight}), and the median scaling relations have higher normalisations, in better agreement with the data and in line with the findings of \citet{Barausse12}.  Although the selection effect is smaller for this specific subsample of galaxies, the model-predicted intrinsic \mbh-$\sigma$ relation is offset slightly above the data, while the predictions for the intrinsic \mbh-\mstar\ relation are slightly below the data.
The effect of including the selection bias on the resolvability of the black hole sphere of influence (middle and bottom panels) is less important than in \figu\ref{fig|ScalingsETGs}, because the small systems for which the sphere of influence is not resolvable tend to live in late-type galaxies in our semi-analytic model. Nevertheless, the selection bias still tends to make the correlations higher in normalisation, slightly flatter, and slightly tighter.

\begin{figure}
    \center{\includegraphics[width=8truecm]{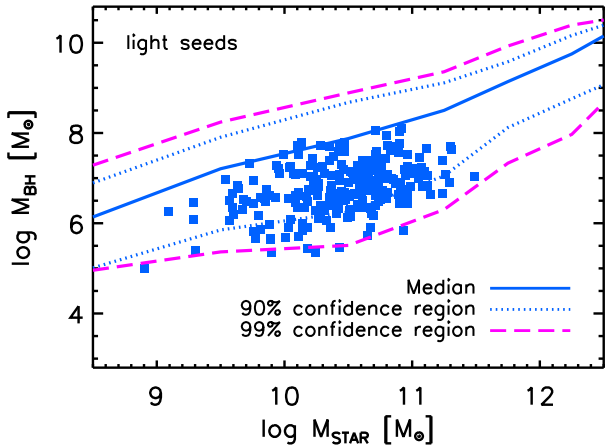}
    \caption{Median black-hole mass as a function of total host-galaxy stellar mass (solid lines) as predicted by the model, for the black holes
with bolometric luminosity  $\log L/{\rm erg\, s^{-1}}>42$. The dotted (long-dashed) lines mark the 90\% (99\%) confidence regions
at given stellar mass, as predicted by our model.
Blue squares are data from \citet{ReinesVolonteri15}. The results are for the light-seed model, but the heavy-seed one gives very similar results.
    \label{fig|ScalingsReines}}
}
\end{figure}

\subsection{The dispersion around the scaling relations: a comparison with observations}
\label{sec:dispersion}

It is important to emphasise that, whatever the exact black-hole seeding recipe adopted in the model, the predicted local (intrinsic) scaling relations reported in \figu\ref{fig|ScalingsLight} for all galaxy types show very large dispersions, especially for galaxy stellar masses $\mstare \lesssim 3\times 10^{11}\, \msune$, with a broad distribution up to two orders of magnitude or more in black hole mass at fixed stellar mass or velocity dispersion. This is because the model tends to retain a significant number of low-mass black holes, which did not accrete much gas along their history (because they live in spirals or satellite galaxies) and which therefore remain closer to their seed masses \citep[e.g.,][]{Volonteri05,Barausse12}. However, for more massive galaxies the dispersion is smaller, with almost no galaxies with $\mstare \gtrsim 5\times 10^{11}\, \msune$ having black holes with $\mbhe \lesssim 10^7\, \msune$.

One interesting way to probe the existence of very low-mass black holes in relatively low-mass galaxies could be to compare with the scaling relations in active galaxies, which are not limited by the spatial resolution issues that heavily affect dynamical measurements of black holes. To this purpose, \figu\ref{fig|ScalingsReines} compares the predictions of our model with the recent sample of 262 broad-line AGNs collected by \citet{ReinesVolonteri15}.  In more detail, the observations are represented by blue squares, while the lines represent the median, the 90\% confidence region (i.e. the 5th and 95th percentiles) and the 99\% confidence region (i.e. the $0.5$th and $99.5$th) of the (model-predicted) \mbh-\mstar\ relation, by assuming a light black-hole seed scenario (the heavy-seed scenario gives very similar results) and considering only systems with bolometric luminosity $\log (L/{\rm erg\, s^{-1}})>42$ \citep[roughly the minimum luminosity probed by][]{ReinesVolonteri15}.
The model's distribution of active black holes has been built by randomly drawing Eddington ratios from a Schechter distribution that extends up to the Eddington limit, in agreement with a number of observations \citep[][]{Kauffmann09,Aird12,Bongiorno12,Schulze15,Jones16}. We have verified that our predicted luminosity function at $z=0$, computed by assuming an average duty cycle of active black holes of 10\% consistent with the results from local surveys \citep[e.g.,][and references therein]{Goulding09,Shankar13,Pardo16}, agrees with the (obscuration-corrected) bolometric luminosity functions of \citet{Hop07} and \citet{SWM}.

First, let us note, as emphasised by \citet{Shankar16BH}, that a lower limit of $L\gtrsim 10^{42}\, \ergse$ should still allow black holes down to a mass of $\mbhe \sim 10^4\, \msune$ to be detected, at least if a non-negligible fraction of these black holes are still accreting at the Eddington limit.  Such low mass black holes do not seem to exist in the \citet[][see also \citealt{Baldassare15}]{ReinesVolonteri15} sample (and in our model, at least in sufficiently large numbers and with high enough Eddington ratios to warrant detection).  Even assuming lower virial factors $f_{\rm vir}$ than those adopted by \citet{ReinesVolonteri15} in deriving black hole masses from their measured FWHMs, as suggested by some groups \citep[e.g.,][and references therein]{Shankar16BH,Yong16}, would not alter these conclusions.

Second, it is clear that the observational sample lies, on average, below the model median predictions, which were tuned to reproduce the data on inactive local galaxies with a significantly higher normalisation.  While the predicted 90\% and 99\% confidence regions for the active black-hole population encompass the data of \citet{ReinesVolonteri15}, the model also predicts the existence of a large number of active higher-mass black holes (above the median), which are not observed. Therefore, either the sample of \citet{ReinesVolonteri15} is biased toward low-luminosity active systems, or our model should be normalised to lower values by a factor $\gtrsim 3$.

The large scatter in our model means that, if the normalisation is decreased, then our model would predict a large tail of very low-mass black holes. Since these are not observed, this would have important consequences for constraining models of the seeds of the supermassive black-hole population.  On the other hand, decreasing the normalisation of the \mbh-\mstar\ relation is by no means straightforward.  While it could be achieved by decreasing black-hole accretion, this would imply a proportional reduction in AGN luminosity, unless a higher radiative efficiency and/or duty cycle are also assumed.  The present calibration of our model already predicts rather high radiative efficiencies/black-hole spins \citep{Sesana14}. Duty cycles of active black holes could instead be constrained by comparing with independent AGN clustering measurements \citep[][and references therein]{Gatti16}. We plan to explore some of these important interrelated issues in future work.

\begin{figure*}
    \center{\includegraphics[width=17truecm]{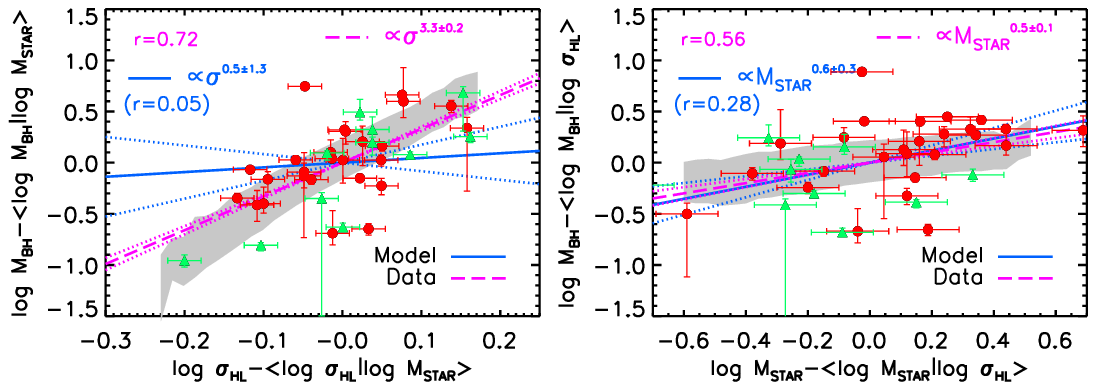}
    \caption{Correlation between the residuals of the \mbh-$\mstare$ relations and those of the $\sigma$-$\mstare$ relation, at fixed stellar mass (left panel), and between the residuals of the \mbh-$\sigma$ relation and those of the $\mstare$-$\sigma$ relation, at fixed velocity dispersion (right panel).  Red circles and green triangles show the ellipticals and lenticulars in the \citet{Savo15} sample. The solid blue and long-dashed magenta lines mark the best fits (for the model's predictions and the observations, respectively), while the dotted lines show the 1$\sigma$ uncertainty on the slope. Also reported are the best fits and the Pearson correlation coefficient $r$. The grey bands show the residuals extracted from the Monte Carlo simulations from \citet{Shankar16BH}, with selection bias on the black-hole gravitational sphere of influence. Note that the correlation of the residuals at fixed stellar mass (left panel) is very strong in the data, but essentially absent in the model.
These results are for the light-seed model (the heavy-seed one yields similar results), selecting only early-type galaxies ($B/T>0.7$) for which the black-hole sphere of influence is resolvable.
    \label{fig|Residuals}}
 }
\end{figure*}

\subsection{Constraints from correlations between residuals}
\label{sec:residuals}
We now compare our model's predictions for the \emph{residuals} of the black hole-galaxy scaling relations to the data. Such correlations are an efficient way of going beyond pairwise correlations between the variables themselves \citep[][]{Bernardi05,ShethBernardi12}. For example, measurements of the \mbh-\sis\ and \mbh-\mstar\ correlations alone do not provide insight about whether \sis\ is more important than \mstar\ in determining \mbh.  This is because the \mbh-\sis\ and \mbh-\mstar\ correlations do not encode complete information about the joint distribution of \mbh, \sis\ and \mstar.  Correlations between the residuals encode some of this extra information.

\begin{figure*}
    \center{\includegraphics[width=17truecm]{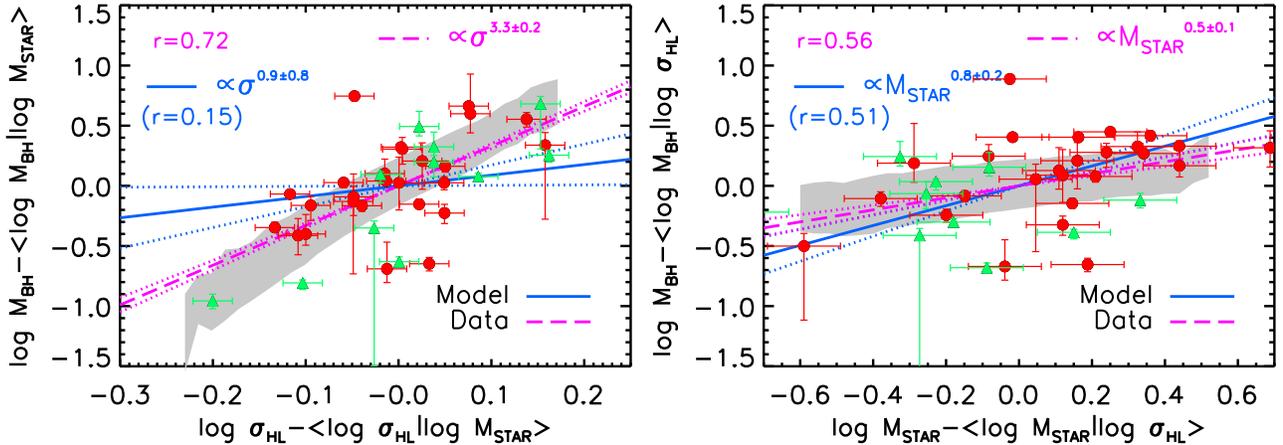}
    \caption{\label{fig|Residuals_horizon}
      Same as \figu\ref{fig|Residuals}, but for the hydrodynamic cosmological simulation Horizon-AGN \citep{dubois,dubois16}. 
However, unlike in \figu\ref{fig|Residuals}, no selection bias on the resolvability of the black-hole sphere of influence has been applied.  Doing so leads to slightly weaker correlations and slightly lower slopes for the fits to the residuals.
Note that the velocity dispersions in the simulation are measured within the effective radius of the galaxy, and are not corrected to the Hyperleda aperture. 
}}
\end{figure*}

\begin{figure}
    \center{\includegraphics[width=8truecm]{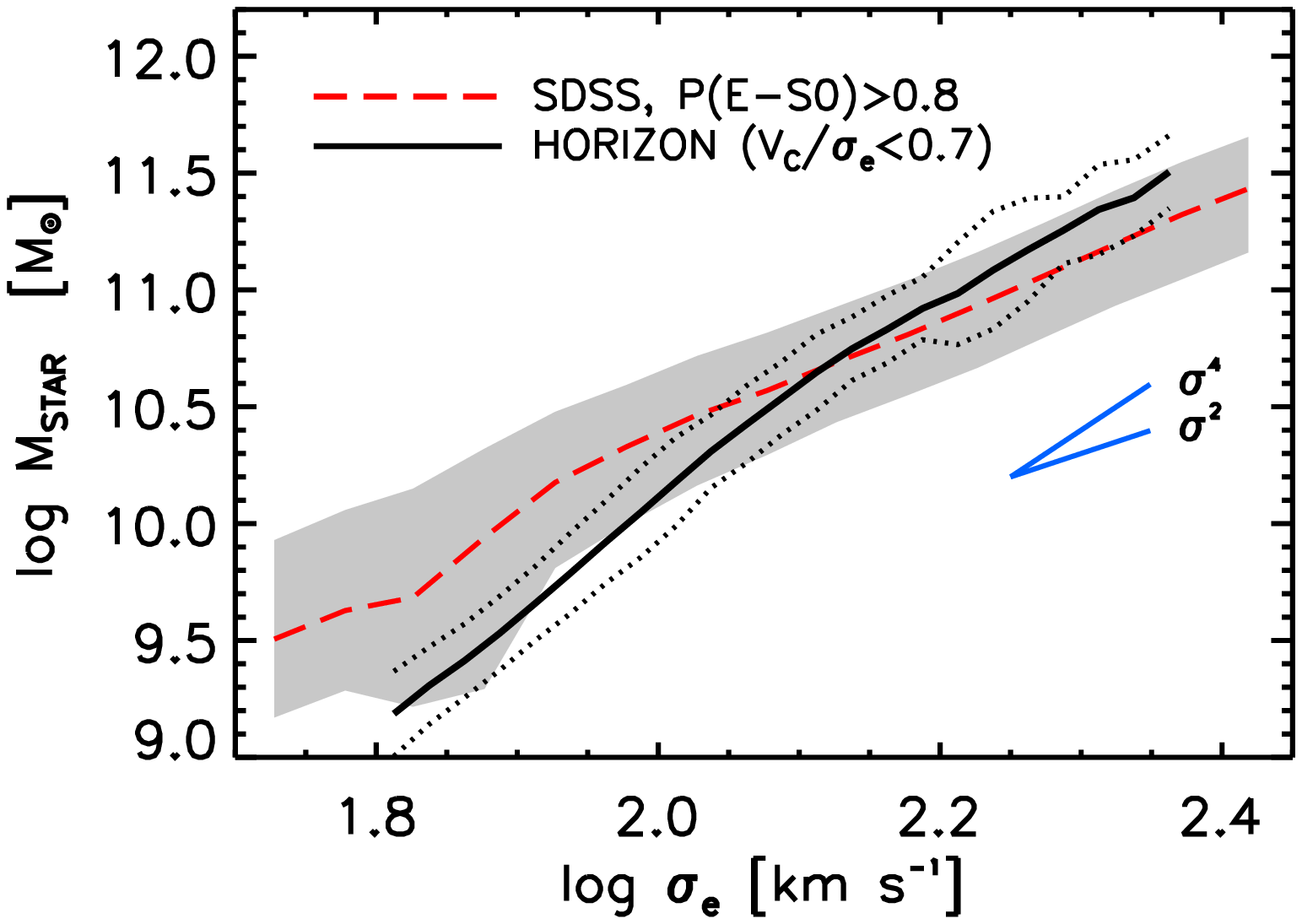}
    \caption{\label{fig|gamma_horizon}
Total stellar mass as a function of velocity dispersion for SDSS galaxies with $P(E-S0)>0.8$ (shaded region), and for the Horizon-AGN simulation \citep[the solid line represents the mean, and the dotted lines represent the 70\% confidence region;][]{dubois,dubois16}. As in the observational sample, we select early-type galaxies from the simulations by only considering systems with $V_c/\sigma<0.7$, where $V_c$ is the rotational velocity. Applying a different cut does not change this figure significantly. Here, for both the SDSS and Horizon-AGN data, $\sigma_e$ is computed within the effective radius. Given the resolution of Horizon-AGN, this quantity is more reliably estimated than the more central velocity dispersion $\sigma_{\rm HL}$ adopted in the previous figures.
}}
\end{figure}

To this purpose, the left and right hand panels of \figu\ref{fig|Residuals} show
$\Delta(\mbhe|\mstare)$ vs $\Delta(\sise|\mstare)$ and
$\Delta(\mbhe|\sise)$ vs $\Delta(\mstare|\sise)$, where
\begin{equation}
\Delta(Y|X)\equiv\log Y-\langle \log Y|\log X \rangle \,
\label{eq|resid}
\end{equation}
is the residual in the $Y$ variable (at fixed $X$) from the log-log-linear fit of $Y(X)$ vs $X$, i.e. $\langle \log Y|\log X \rangle$.
The magenta long-dashed and dotted lines in each panel show the best fit and 1$\sigma$ uncertainties on the correlations between residuals in the \citet{Savo15} dataset:  red circles and green triangles represent ellipticals and lenticulars.  We obtained the magenta lines by running 200 iterations following the steps outlined in \citet{shankar_new}, which include errors in both variables. At each iteration we eliminate three random objects from the original sample. From the full ensemble of realizations, we measure the mean slope and its 1$\sigma$ uncertainty.

The blue solid and dotted lines show a similar analysis in our semi-analytic model.  However, in this case, we randomly produce 30 mock samples of $\sim 75$ galaxies each, with $B/T>0.7$ and resolvable black-hole spheres of influence. From the full ensemble of mock realizations, we then extract the mean slope and Pearson coefficient.

The correlations in \figu\ref{fig|Residuals} show that, in the data, the  velocity dispersion is more strongly correlated with the black-hole mass, with a mean Pearson coefficient of $r=0.72$, than stellar mass, for which the Pearson coefficient is $r=0.56$. The model instead predicts just the opposite, with almost zero correlation with velocity dispersion (mean $r=0.05$), but with a correlation with stellar mass consistent with the data, though still rather weak (mean $r=0.28$). It is nevertheless important to realise that even an intrinsically weak correlation with velocity dispersion at fixed stellar mass does not necessarily imply that the \emph{total} correlation with velocity dispersion is small. In fact, following Appendix B in \citet{shankar_new}, the total dependence of the black-hole mass on velocity dispersion can be summarised as $\mbhe\propto\sigma^{\beta}\mstare^{\alpha}\propto\sigma^{\beta+\alpha\,\gamma}$, where $\gamma$ comes from $\mstare\propto\sigma^{\gamma}$ (where we have ignored any other explicit dependence on, e.g., S\'{e}rsic index or galaxy size). Since the model predicts $\gamma\approx 4$, and the residuals in \figu\ref{fig|Residuals} (solid blue lines) yield $\beta\sim 0.5$ and $\alpha\sim 0.6$,
one obtains a total dependence of $\mbhe\propto\sigma^{3}$, consistent with \figu\ref{fig|ScalingsETGs} (middle left panel).
Also note that a slope $\gamma\approx 4$ for the \mstar-\sis\ relation is in itself already in tension with the ($V_{\max}$-corrected) SDSS observations, which rather
prefer a slope $\gamma\approx2.2$--2.5, at least for $\mstare\gtrsim2\times10^{10} M_\odot$ (c.f. also \figu\ref{fig|gamma_horizon} below).

For completeness, \figu\ref{fig|Residuals} also shows results from the Monte Carlo simulations performed by \citet[][grey bands]{Shankar16BH}, which assume an intrinsic correlation of the type $\mbhe \propto \sise^{4.5} \mstare^{0.5}$ and include the selection bias on $r_{\rm infl}$. These models reproduce the observed trends.

Several comments are in order.
First, our results are robust against the efficiency of the AGN feedback. Indeed, in our reference model we fixed the AGN feedback efficiency to obtain a stellar mass function as close as possible to the observations at the high-mass end (c.f. \figu\ref{fig|ScalingsMstar}, right panel). By slightly decreasing this efficiency (e.g. by a factor $\sim 3$), the agreement with the observed stellar mass function in \figu\ref{fig|ScalingsMstar} slightly worsens in the high-mass end, while the \mbh-\sis\ relation steepens\footnote{Note that, in our model, increasing the AGN feedback efficiency \emph{decreases} the slope of the \mbh-\sis\ relation, mainly because AGN feedback is more effective at clearing the gas in more massive galactic hosts, thus inhibiting the growth of especially the more massive black holes.} ($\mbhe\propto \sise^{3.4}$ after including the effect of the selection bias), but not enough to fully match the observations \citep[for which slopes  $\sim 4.5-5$ are usually quoted in the literature, e.g.][]{Graham11,KormendyHo}. More importantly, the correlations of the residuals remain almost unchanged ($r=0.1$ when fixing \mstar, and $r=0.3$ when fixing \sis). Second, we have checked that these correlations are not improved by considering a different choice of $k$ in \eq\ref{eq|rinfl};
by assuming no bias on the resolvability of the black-hole sphere of influence;
by considering all galaxies rather than just bulge-dominated ones;
by using the bulge mass instead of the total stellar mass;
or by using the the SDSS $\sigma$-$\mstare$ relation \citep[e.g., by the fits of][]{Sesana16} to compute velocity dispersions.
Third, the model's residuals shown in \figu\ref{fig|Residuals} do not account for  measurement errors (i.e. we compute the residuals for the model's exact predictions, without folding in any observational uncertainties). We have verified that including these errors can yield a steeper slope $\beta \sim 2$ for the residuals at fixed \mstar, but this does not strengthen the correlations of the residuals shown above (essentially because the error on $\beta$ also grows when $\beta$ grows).

To check if our results are due to the particular implementation of black-hole accretion and AGN feedback in our semi-analytic model, and/or lack of sufficient ``coupling'' between velocity dispersion and AGN feedback, we have performed a similar analysis of the Horizon-AGN simulation \citep{dubois,dubois16}.  This is a hydrodynamic cosmological simulation of a box with size $100$ Mpc$/h$, run with the adaptive mesh refinement code RAMSES~\citep{ramses}, with $10^9$ dark-matter particles and a minimum mesh size of 1 kpc.  The simulation includes gas cooling, star formation, feedback from stars and AGNs (see~\citealp{dubois} for the details of the numerical modelling and~\citealp{2016MNRAS.460.2979V} for the discussion of the correlations between galaxies and black holes in Horizon-AGN). Galaxies are extracted with a galaxy finder running on star particles. Since the simulation assumes a Salpeter \citep{SalpeterIMF} IMF, we have reduced galaxy stellar masses by 0.25 dex \citep[e.g.,][]{Bernardi10} to match the Chabrier IMF~\citep{Chabrier03} adopted in this work. The velocity dispersion is measured from its components along each direction of the cylindrical coordinates oriented along the galaxy's spin axis (i.e. $\sigma_r$, $\sigma_t$ and $\sigma_z$ for the radial-, tangential-, and $z$-component respectively), thus $\sigma^2=(\sigma_r^2+\sigma_t^2+\sigma_z^2)/3$. The velocity dispersion  of each galaxy is measured using only star particles within the effective radius of the galaxy.  The results are shown in \figus\ref{fig|Residuals_horizon} and \ref{fig|gamma_horizon}.  They are in qualitative agreement with those obtained with our semi-analytic model. In more detail, Horizon-AGN also predicts $\gamma\approx 4$, in tension with the data, and a very weak correlation between the residuals when fixing \mstar. Also note that in \figu\ref{fig|Residuals_horizon} we have not applied any selection bias (unlike in the case of the semi-analytic model in \figu\ref{fig|Residuals}). Restricting to systems with resolvable spheres of influence actually makes the correlations in the residuals of the Horizon-AGN simulation even weaker.

We plan to consider different (and possibly stronger) models of AGN feedback in future work.  For now, the discrepancies highlighted in \figus\ref{fig|Residuals} and \ref{fig|Residuals_horizon} clearly show that correlations between the residuals of the \mbh-\sis\ and \mbh-\mstar\ scaling relations are more powerful than the scaling relations themselves at constraining models for the co-evolution of black holes and their host galaxies.

\section{Discussion}
\label{sec|discu}

Concerning the normalisation of the black-hole scaling relations, at face value our model clearly fails to reproduce the observed \mbh-\sis\ and \mbh-\mstar\ relations at the same time.
Without invoking any selection effect, one could attempt to improve the simultaneous match to both the observed scaling relations by fine-tuning some of the key parameters in the model, namely the one controlling gas accretion onto the black hole and/or the energetic feedback from the central active nucleus. For example, increasing the efficiency of AGN feedback could in principle decrease the stellar masses of the host galaxies at fixed black hole mass, thus possibly improving the match to the \mbh-\mstar\ relation (top right panel of \figu\ref{fig|ScalingsLight}). However, this would then spoil the good match to the velocity dispersion-stellar mass relation.
Similarly, one could increase the accretion onto the central black hole at fixed star formation rate and final stellar mass of the host galaxy, with the aim of improving the match to the \mbh-\mstar\ relation. This however would proportionally increase the \mbh-\sis\ relation above the data, when velocity dispersions are chosen to faithfully track those from the observed SDSS $\sigma-\mstare$ relation (magenta lines in the left panels of \figu\ref{fig|ScalingsLight}).
Concerning the dispersion around the scaling relations, our semi-analytic model, which self-consistently evolves black holes from seeds at high redshifts,
naturally predicts very broad distributions in black-hole mass at fixed velocity dispersion or stellar mass, at relatively low masses $\mstare \lesssim 3\times 10^{11}\, \msune$.

Overall, these effects are in line with the results of the Monte Carlo simulations presented in \citet{Shankar16BH}, though the intrinsic scatters
assumed there, especially in the \mbh-\sis\ relation, were always below $0.3$ dex. This was needed to avoid too flat slopes in the ``observed'', biased relations.
Indeed, here our predicted slope of the \mbh-\sis\ relation, inclusive of observational bias (e.g., middle and bottom left panels of \figu\ref{fig|ScalingsLight}), is approximately $\mbhe\propto\sigma^{\beta}$ with slope $\beta\sim 3-3.5$, significantly flatter than the slopes $\beta \gtrsim 4.5-5$ usually quoted in the literature \citep[e.g.,][]{Graham11,KormendyHo}.

Conversely, a key point of our present work is that, irrespective of the chosen masses for the seed black holes, the model predicts relatively tight scaling relations at high stellar masses $\mstare \gtrsim 5\times 10^{11}\, \msune$. In this respect, our model does not support the conjecture put forward by \citet{Batcheldor07}, according to which the \mbh-\sis\ relation is only an upper limit of a more or less uniform distribution of black holes. This is also in line with the Monte Carlo tests performed by \citet[][their \figu11]{Shankar16BH}, in which a very broad distribution of black holes extending to the lowest masses in all types of galaxies is highly disfavoured.
Nevertheless, \figu\ref{fig|ScalingsReines} shows that the present data on active galaxies may still be consistent with our model,
once the proper flux limits and Eddington ratio distributions are accounted for.
Therefore, an intrinsically broad distribution for the black-hole mass in relatively small galaxies with mass $\mstare \lesssim 3\times 10^{11}\, \msune$, extending down to black holes of mass $\mbhe \gtrsim 10^5\, \msune$ or so, cannot be excluded by present data, though it is highly disfavoured in more massive galaxies.
Models with such broad distributions, however, still tend to produce scaling relations flatter than observed, once the bias on the black-hole sphere of influence and on galaxy morphology is folded in the model predictions, as shown in \figus\ref{fig|ScalingsLight} and \ref{fig|ScalingsETGs}.

Finally, we stress again that our semi-analytic model fails to reproduce the observations when it comes to correlations between the \emph{residuals} of the scaling relations. In more detail, the model is consistent with the data for
the correlation between  the residuals in the \mbh-$\sigma$ relation and those in the $\mstare$-$\sigma$ relation, at fixed velocity dispersion. However, the model predicts almost no correlation between the residuals of the \mbh-$\mstare$ relations and those of the $\sigma$-$\mstare$ relation, at fixed stellar mass, while the data hint at a rather strong correlation. We have verified that these results are robust against changing the parameters of the model (namely the AGN feedback efficiency and the parameter regulating black-hole accretion). Moreover, we have  shown that the same weak correlation between the residuals at fixed stellar mass is also obtained in the hydrodynamic cosmological simulation Horizon-AGN~\citep{dubois,dubois16}, which includes thermal quasar-mode feedback and jet-structured radio-mode AGN feedback (i.e., processes that are expected to induce a stronger coupling between black-hole mass and velocity dispersion).
Another noteworthy point is that both our model and Horizon-AGN tend to produce slopes for the \sis-\mstar\ relation that are significantly steeper than the data, although in Horizon-AGN numerical resolution effects may bias the measurements of the velocity dispersion in lower-mass galaxies \citep[see discussion in][]{dubois16}.

Our interpretation is that the weak correlation of the residuals at fixed stellar mass may be at least partly due to current AGN feedback models being possibly too weak to capture the full effect of the black hole on the \emph{stellar} velocity dispersion.
While this is expected in a semi-analytic model such as ours, where the AGN feedback is typically assumed to simply eject gas from the nuclear region~\citep{Granato04,Barausse12}, it is more surprising in the Horizon-AGN simulation, which captures the back-reaction of the ejected gas onto the stellar and dark-matter dynamics~\citep{peirani16}.
Nonetheless, in hydrodynamic cosmological simulations such as Horizon-AGN, the spatial resolution is 1 kpc at best and may limit our capability to properly capture the interaction of AGN winds with gas and its impact on the dynamics within galactic bulges.
Moreover, the Horizon-AGN simulation currently employs a rather crude model of AGN feedback, mostly based on simple gas heating,
to mimic the so-called ``quasar-mode'' feedback. More realistic models of AGN feedback, also inclusive of
momentum-driven winds and stronger coupling with the surrounding interstellar medium \citep[e.g.,][]{bieri17}, might possibly be more effective at improving the comparison with
the black hole-galaxy scaling relations and their residuals. We should also mention that another possibility that has been put forward in the literature is
that AGN feedback may \emph{not} be the main cause of the scaling relations, which might be ascribed instead to a common gas supply for the galaxy and the black hole, regulated by gravitational torques~\citep{2017MNRAS.464.2840A}.

\section{Conclusions}
\label{sec|conclu}

We have compared  the predictions of a comprehensive semi-analytic model of galaxy formation, which self-consistently evolves supermassive black holes from high-redshift seeds by accounting for gas accretion, mergers and AGN feedback,
with the observed scaling relations of the masses of supermassive black holes with stellar mass and velocity dispersion. Our main conclusions are:

\begin{enumerate}

\item At $\mstare \gtrsim 5\times 10^{11}\, \msune$, the dispersion in black-hole mass at fixed stellar mass is $\lesssim1$~dex -- very few black holes with masses $\mbhe \lesssim 10^{7}\, \msune$ are predicted in such massive galaxies.  However, for galaxies having $\mstare \lesssim 3\times 10^{11}\, \msune$, the distribution of \mbh\ at fixed $\sigma$ or \mstar\ is broad.

\item  Observational selection effects, associated with resolving the black-hole sphere of influence and/or with selecting bulge-dominated/elliptical galaxies, tighten the \mbh-\sis\ and \mbh-\mstar\ scaling relations, bringing them into better agreement with the observations.

\item No evident variation in AGN feedback and/or black-hole accretion efficiencies can provide a \emph{simultaneous} match to both scaling relations.  This supports previous work suggesting an internal inconsistency between the observed \mbh-\sis\ and \mbh-\mstar\ relations.

\item Galaxy-evolution models (our semi-analytic one as well as the Horizon-AGN simulation) predict almost \emph{no} correlation between the residuals of the \mbh-$\mstare$ relations and those of the $\sigma$-$\mstare$ relation, at fixed stellar mass.
  Since the data hint at a rather strong correlation, this calls for revamped AGN feedback recipes in the next generation of cosmological galaxy-evolution models, or for a re-assessment
of the importance of gravitational torques in regulating the black hole-galaxy co-evolution~\citep{2017MNRAS.464.2840A}.

\end{enumerate}

\section*{Acknowledgments}
We thank M. Volonteri for insightful conversations. EB acknowledges support from the H2020-MSCA-RISE-2015 Grant No. StronGrHEP-690904 and from the APACHE grant (ANR-16-CE31-0001) of the French Agence Nationale de la Recherche. 
This work has made use of the Horizon Cluster, hosted by the Institut d'Astrophysique de Paris. We thank Stephane Rouberol for running this cluster so smoothly.

\bibliographystyle{mn2e_Daly}

\label{lastpage}
\end{document}